\documentclass[prb,twocolumn,showpacs,superscriptaddress]{revtex4}

\usepackage{graphicx}
\usepackage{amsmath}
\usepackage{color}
\usepackage{amssymb}
\usepackage{bm}

\newcommand{\state}[3]{$|#1 #2 \!\! #3\rangle$}
\newcommand{\dstate}[6]{$|#1 #2 \!\! #3 #4 #5 \!\! #6\rangle$}

\newcommand{\edag}[1]{a_{c#1}^{\dagger}}
\newcommand{\e}[1]{a_{c#1}^{\phantom \dagger}}
\newcommand{\hdag}[1]{a_{v#1}^{\dagger}}
\newcommand{\h}[1]{a_{v#1}^{\phantom \dagger}}
\newcommand{\cdag}[1]{c_{\textbf{#1}}^{\dagger}}
\newcommand{\can}[1]{c_{\textbf{#1}}^{\phantom \dagger}}
\newcommand{\pDE}[0]{p_{\text{DE}}}
\newcommand{\pII}[0]{p_{\text{II}}}

\newcommand{\ua}{\uparrow}
\newcommand{\da}{\downarrow}

\newcommand{\hzwei}{\frac{2}{\hbar}}
\newcommand{\imag}[1]{\hspace{1mm}\Im\big(#1\big)}

\begin{document}

\title{Ultrafast Dynamics of Carrier Multiplication in Quantum Dots}

\author{Franz Schulze}
\email[]{schulze@itp.tu-berlin.de}
\author{Mario Schoth}
\affiliation{Institut f\"ur Theoretische Physik, Nichtlineare Optik und Quantenelektronik, Technische Universit\"at Berlin,
Hardenbergstr. 36, 10623 Berlin, Germany}
\author{Ulrike Woggon}
\affiliation{Institut f\"ur Optik und Atomare Physik, Nichtlineare Optik, Technische Universit\"at Berlin, Stra\ss e des 17. Juni 135,
10623
Berlin, Germany}
\author{Carsten Weber}
\author{Andreas Knorr}
\affiliation{Institut f\"ur Theoretische Physik, Nichtlineare Optik und Quantenelektronik, Technische Universit\"at Berlin,
Hardenbergstr. 36, 10623 Berlin, Germany}

\date{May 27, 2011}

\begin{abstract}
A quantum-kinetic approach to the ultrafast dynamics of carrier multiplication in semiconductor
quantum dots is presented. We investigate the underlying dynamics in
the electronic subband occupations and the time-resolved optical emission spectrum, focusing on the interplay between the light-matter and the Coulomb interaction.
We find a transition between qualitatively differing behaviors of carrier multiplication,
which is controlled by the ratio of the interaction induced time scale and the pulse duration of the
exciting light pulse. On short time scales, i.e., before intra-band relaxation, this opens the possibility of detecting carrier
multiplication without refering to measurements of (multi-)exciton lifetimes.
\end{abstract}


\maketitle

\section{Introduction}

Typically, in a semiconductor excited by a single photon, the
conversion efficiency of light energy into electrical energy is strongly limited by the fact that per photon only a single
electron-hole pair (or exciton) can be created. For single-junction solar
cells, this determines the well-known Shockley-Queisser
limit.\cite{ShockleyQueisser}
Using multiple materials with different band gaps (multi-junction cells), a larger part of the solar
spectrum can be covered, and thus the conversion efficiency can be increased
beyond this limit.
Another way is to use materials which exhibit the creation of multiple excitons per incident photon (multiple-exciton generation -
MEG), thus increasing the quantum yield for incident photons with energies higher than twice the band gap.
For this reason, carrier multiplication is currently the subject of intensive
investigation in nanostructures such as
graphene structures\cite{Graphene2010, Winzer2010}
and
semiconductor quantum dots (QDs).\cite{Califano2009,McGuire2010,Tyagi2011,Franceschetti2006,Delerue2010,Shabaev2006,ScreeningZunger,
Nair2008,Pijpers2007,Schaller2004,Witzel2010}

Most of the experimental investigations focus on the decay dynamics of the carrier
population in the conduction
band as a measure of MEG: While an exciton decays radiatively on a time scale of tens to hundreds of picoseconds,
multiexcitations can decay much faster (on a picosecond time scale) due to additional Coulomb-mediated impact ionization and Auger
scattering channels. Thus, the detection of different decay time constants in dependence on the incident light frequency
is used as an indication and quantitative measure of MEG.
One disadvantage of these approaches is the ambiguity accompanying the discrimination of MEG from
other effects in quantum dots such as charged excitons.\cite{Califano2009,McGuire2010,Tyagi2011}

Corresponding to experimental investigations, theory has focused on calculations of (multi-)exciton decay rates which were
found to be crucial for
MEG and its efficiency in concrete material systems.\cite{Franceschetti2006,Delerue2010}
The theoretical descriptions are typically restricted to Bloch equations including phenomenological scattering rates as well as pseudo-potential
approaches without quantum-kinetic time resolution of the carrier multiplication process. The time dynamics of absorption
bleaching was investigated by Shabaev et al. in Ref.~\onlinecite{Shabaev2006}.

Here, we present a quantum-kinetic approach to carrier multiplication in a semiconductor QD including the time dynamics of
the relevant Coulomb processes (impact ionization and Auger
recombination), studied in the electronic occupations and the
time-resolved optical emission spectrum. We propose that distinct quantum-kinetic signatures of the carrier dynamics due to different dynamical
paths can be used as an experimental verification of carrier multiplication via the time-resolved optical emission
in quantum dot structures: This proposal is, in principle, advantageous compared to measurements of exciton or multiexciton
decay times due to a clear interpretation of the signal.

\section{\label{model}Model System}

\begin{figure}[htb]
\includegraphics[width=80mm,clip]{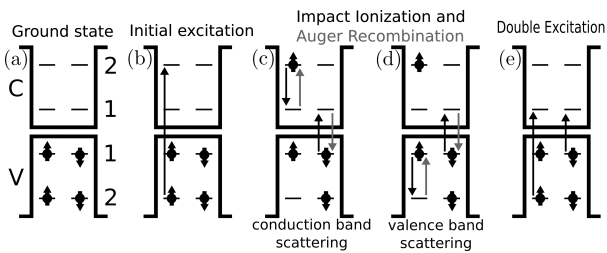}
\caption{(a) Quantum dot model in its ground state consisting of eight electronic states [four
in the conduction (c) and four in the
valence band (v)] exhibiting spin degeneracy. (b) Creation of an electron density in the state \state{c}{2}{\uparrow}
by a classical light pulse. (c) Impact ionization (black) and Auger recombination (grey) due to
energy-conserving conduction band electron-electron scattering.
(d) Impact ionization (black) and Auger recombination (grey) due to energy-conserving electron-electron scattering in
the valence band. (e) Simultaneous excitation of two electrons from the valence to the conduction band.}
\label{modelsystem}
\end{figure}

The QD model underlying this work consists of eight electronic states exhibiting spin degeneracy, i.e., four in
the conduction (c) and four in the valence band (v), depicted schematically in Fig.~\ref{modelsystem}(a).
The conduction band states are labeled by $|c {\lambda s}\rangle$,
while the valence band states are labeled by $|v {\lambda s}\rangle$; here, $\lambda$ denotes the subband index
(two subband states for each band) and $s$ the spin index.

The full system Hamiltonian reads
\begin{equation}
H = H_{0} + H_{\text{cl}} + H_{\text{Coul}} ,
\end{equation}
where $H_{0}$ represents the free kinetics of the electrons:
\begin{equation}
H_{0} = \sum_{\stackrel{\lambda=1,2}{s=\uparrow,\downarrow}} \left( \epsilon^c_{\lambda s} \edag{\lambda s} \e{\lambda s} +
\epsilon^v_{\lambda s} \hdag{\lambda s} \h{\lambda s} \right).
\end{equation}
$\edag{\lambda s}$ ($\hdag{\lambda s}$) and $\e{\lambda s}$ ($\h{\lambda s}$) are the creation and annihilation operators of an
electron in the conduction (valence) band and subband $\lambda$ with spin
$s$ and energy $\epsilon^{c(v)}_{\lambda s}$, respectively.

Initially, the systems is excited by a classical light pulse close to the transition \state{v}{2}{\uparrow}
$\rightarrow$\state{c}{2}{\uparrow} [Fig.~\ref{modelsystem}(b)], creating the
single-exciton state \dstate{c}{2}{\ua}{v}{2}{\ua}: This interaction between the electrons and the incident light pulse is
treated semiclassically and in dipole and rotating wave approximation:
\begin{equation}
H_{\text{cl}} = M_{2 \uparrow 2\uparrow}^{cv} E(t) \hspace{1.2mm} \edag{2 \uparrow} \h{2 \uparrow} + \text{h.a.  ,}
\end{equation}
where 
$E(t)$ is the classical light field at the QD location, taken to be a Gaussian pulse $E(t) = \tilde{E}(t) \cos(\omega_L t) \text{ with
} \tilde{E}(t) = E_0 \exp\{-t^2/\tau^2\}$, with the pulse width $\tau$. The pulse area is defined
via the envelope of the pulse, $\theta = \int_{-\infty}^\infty dt' M_{2\uparrow 2\uparrow}^{c v} \tilde{E}(t')/\hbar$. $M_{2\uparrow
2\uparrow}^{cv}$ is the dipole matrix element of the considered transition.

To focus on the underlying dynamics of MEG, the full
carrier-carrier interaction Hamiltonian is restricted for large
parts of the article to the impact ionization (II) and Auger recombination
(AR) processes of conduction band electrons, depicted in Fig.~\ref{modelsystem}(c).
The additional consideration of II and AR of valence band electrons
[Fig.~\ref{modelsystem}(d)], which occurs on an equal footing, complicates the
presentation of the underlying physics while it does not
qualitatively change the observed phenomena. For completeness,
we include calculations of both scattering processes at the end
of the numerical discussion including a comparison
between the system dynamics excluding and including II and AR processes of
valence band electrons.

The impact ionization process
annihilates electrons in the states \state{c}{2}{\uparrow} and \state{v}{1}{\downarrow} and creates electrons in
the multi-carrier state \dstate{c}{1}{\uparrow}{c}{1}{\downarrow} [black arrows in Fig.~\ref{modelsystem}(c)], thus describing the
transition between the single-exciton state \dstate{c}{2}{\ua}{v}{2}{\ua} and the two-exciton state $|c 1 \!\! \ua c 1 \!\! \da v 1
\!\! \da v 2 \!\! \ua\rangle$.
The Auger recombination as the inverse process is described by the Hermitian adjoint [grey arrows in
Fig.~\ref{modelsystem}(c)].
The corresponding carrier-carrier interaction Hamiltonian reads
\begin{equation}
H_{\text{Coul}} =
V^{\text{II}}
\hspace{1.2mm} \edag{1 \uparrow} \edag{1 \downarrow} \h{1 \downarrow} \e{2 \uparrow}
+ V^{\text{AR}}
\hspace{1.2mm} \edag{2 \uparrow} \hdag{1 \downarrow} \e{1 \downarrow} \e{1 \uparrow} ,
\end{equation}
with $V^{\text{II}} = V_{1 \uparrow 1 \downarrow 1 \downarrow 2 \uparrow}^{c \hspace{1.4mm} c \hspace{1.4mm} v \hspace{1.4mm} c} =
(V^{\text{AR}})^*$. The Coulomb coupling elements $V^{\text{II}}$ and $V^{AR}$ are taken to be real and are thus equal,
$V^{\text{II}} = V^{\text{AR}} \equiv V$.

The theoretical description developed in this work is inherently non-Markovian and hence allows
to study the dynamics including
non-energy-conserving scattering processes.
This will be utilized later to study the effect of off-resonant Coulomb scattering on the efficiency of the impact ionization
process.

For our calculations, we assume that the coupling elements are determined by an ab initio theory such as in
Ref.~\onlinecite{Kang1997}, thus focusing on the ultrafast time
scale of the quantum-kinetic dynamics. The parameters are chosen for the PbSe QD of
Ref.~\onlinecite{Shabaev2006}, yielding approximate values for the energetic and coupling parameters.

\section{\label{electronicSystem}Dynamical equations}

The quantum-kinetic dynamics is derived within an equation of motion approach.\cite{Rossi2002}
The finite character of the four-electron system is used to close the hierarchy of expectation values arising from
the many-particle interaction
Hamiltonian $H_{\text{Coul}}$, making the calculated results essentially non-perturbative with respect to the Coulomb
interaction.
The restriction to four electrons is a good assumption if the interaction between the electrons inside the quantum dot and
the electrons of the
surrounding substrate is negligibly weak. \cite{Dachner:PhysStatusSolidiB:10} Within this assumption, all correlations containing ten
or more
electron operators vanish because they describe correlations between more than four electrons.

Since we focus on ultrashort time scales (subpicosecond), the influence of radiative decay on the carrier dynamics, which acts
on comparibly long time scales ($T_1 \geq 1$ ns) resulting from the small energy gaps, will be
neglected throughout this work. 
However, pure dephasing processes are considered via a
phenomenological constant $\gamma_{\text{PD}}$. 

It turns out that for the system introduced in Sec.~\ref{model} including the
impact ionization and Auger recombination of only the conduction
band electrons, a reduction to a two-electron description yields results close to the full calculations involving all
electron correlations.
Therefore, to achieve greater clarity, we will
use the two-electron level for a discussion of the system properties. The final dynamical results presented in the figures,
however, are
calculated using the full set of equations, cf. the appendix for the equations. We would like to note that an analogous set of
dynamical equations was used in Ref.~\onlinecite{Shabaev2006} to discuss the time dynamics of absorption bleaching in QDs.

To present a clear depiction of the processes and their mutual dependencies, the operator abbreviations used are listed in
table~\ref{operator_abbreviations}.
\begin{table}[htb]
\begin{tabular}{c c c}
\hline
\hline
Abbreviation &\hspace*{0cm} &Operator\\
\hline
\hline
$f_{\lambda s}^c $ && $\edag{\lambda s} \e{\lambda s}$\\
$f_{\lambda s}^v $ && $\hdag{\lambda s} \h{\lambda s}$\\
$p_{2\uparrow}$ && $\edag{2\uparrow} \h{2\uparrow}$\\
$p_{\text{DE}}$ && $\edag{1\uparrow} \edag{1\downarrow} \h{1\downarrow} \h{2\uparrow}$ \\
$p_{\text{II}}$ && $\edag{1\uparrow} \edag{1\downarrow} \h{1\downarrow} \e{2\uparrow}$\\
\hline
\hline
\end{tabular}
\caption{Operator abbreviations: $f_{\lambda s}^{c(v)}$ is the occupation operator in subband $\lambda$ with spin
$s$ in the conduction (valence) band. $p_{2\uparrow}$ describes the transition of an electron from state \state{v}{2}{\uparrow} to
\state{c}{2}{\uparrow} [cf. Fig.~\ref{modelsystem}(b)]. The operator of the double excitation $p_{\text{DE}}$ [cf.
Fig.~\ref{modelsystem}(e)] annihilates two electrons in the states \state{v}{2}{\uparrow} and
\state{v}{1}{\downarrow} and creates two
electrons in the states \state{c}{1}{\uparrow} and \state{c}{1}{\downarrow}. The operator of impact ionization
$p_{\text{II}}$ [cf.
Fig.~\ref{modelsystem}(c)] describes the correlated transition between the states \state{v}{1}{\downarrow},
\state{c}{1}{\downarrow} and the states \state{c}{2}{\uparrow}, \state{c}{1}{\uparrow}.}
\label{operator_abbreviations}
\end{table}

\subsection{\label{linear_absorption}Energetic Structure and absorption of the QD}

Before we investigate the full dynamics, we first characterize the electronic system via the linear optical
absorption:
\begin{equation}
\alpha(\omega) \sim \omega \Im \left( \frac{P(\omega)}{\epsilon_0 E(\omega)} \right).
\end{equation}
Since we are interested in the energetic structure of the system under excitation of the transition
$|v 2 \uparrow \rangle \rightarrow |c 2 \uparrow \rangle$, we can restrict to the microscopic polarization
$\langle p_{2\uparrow} \rangle$ [cf. table~\ref{operator_abbreviations}], which determines the macroscopic polarization
$P(t)$ via $P(t) \sim M^{cv}_{2\uparrow 2\uparrow} \langle p_{2\uparrow} \rangle(t) + c.c.$

The system of dynamical equations defining the linear absorption $\alpha(\omega)$ via $\langle p_{2\uparrow} \rangle$
reads:
\begin{align}
d_t \langle p_{2\uparrow} \rangle = &\left[ \frac{i}{\hbar} \left( \epsilon_{2\uparrow}^c - \epsilon_{2\uparrow}^v \right) - \gamma_{\text{PD}} \right] \langle p_{2\uparrow} \rangle \nonumber \\
&+ \frac{i}{\hbar} V \langle p_{\text{DE}} \rangle + \frac{i}{\hbar} E(t) M^{cv}_{2\uparrow 2\uparrow} ,
\end{align}
\begin{align}
d_t \langle p_{\text{DE}} \rangle = &\left\{ \frac{i}{\hbar} \left[\left(\epsilon_{1\uparrow}^c -
\epsilon_{2\uparrow}^v\right) + \left(\epsilon_{1\downarrow}^c - \epsilon_{1\downarrow}^v \right)\right] - \gamma_{\text{PD}}
\right\} \langle p_{\text{DE}} \rangle \nonumber \\
&+ \frac{i}{\hbar} V \langle p_{2\uparrow} \rangle.
\end{align}

The analytical solution of the absorption spectrum can be obtained via Fourier transformation, yielding
\begin{equation}
\alpha(\omega) \sim \Im \left( \frac{\omega |M^{cv}_{2\uparrow 2\uparrow}|^2}{
\Delta\varepsilon - i \hbar \gamma_{\text{PD}} - \frac{V^2}{\hbar \omega - \left[\left( \epsilon_{1\uparrow}^c - \epsilon^v_{2\uparrow} \right) + \left( \epsilon_{1\downarrow}^c - \epsilon_{1\downarrow}^v\right)\right]  - i \hbar \gamma_{\text{PD}}}} \right) 
\label{EQ:Absorption} .
\end{equation}
It can be seen that for a vanishing Coulomb coupling, the real part of the denominator vanishes for an energy detuning
$\Delta \varepsilon = \hbar \omega - (\epsilon_{2\uparrow}^c - \epsilon^v_{2\uparrow}) = 0$ between the incident light
field and the
single-exciton energy $(\epsilon_{2\uparrow}^c -
\epsilon^v_{2\uparrow})$.
For non-vanishing and sufficiently large Coulomb coupling, the detuning
$\Delta = (\epsilon_{2\uparrow}^c - \epsilon^v_{2\uparrow})
- [( \epsilon_{1\uparrow}^c - \epsilon^v_{2\uparrow}) + (\epsilon_{1\downarrow}^c -
\epsilon_{1\downarrow}^v )]$
between the energy of the
single-exciton generated by the light field [Fig.~\ref{modelsystem}(b)] $(\epsilon_{2\uparrow}^c -
\epsilon^v_{2\uparrow})$
and the energy of the two-exciton generated by the double excitation $\langle \pDE \rangle$ or impact ionization
$\langle \pII \rangle$ [Fig.~\ref{modelsystem}(c),(e)]
$ [( \epsilon_{1\uparrow}^c - \epsilon^v_{2\uparrow}) + (\epsilon_{1\downarrow}^c -
\epsilon_{1\downarrow}^v )]$ dominates the root
of the denominator and therefore the maxima of the absorption.

\begin{figure}[htb]
\includegraphics[width=86mm,clip]{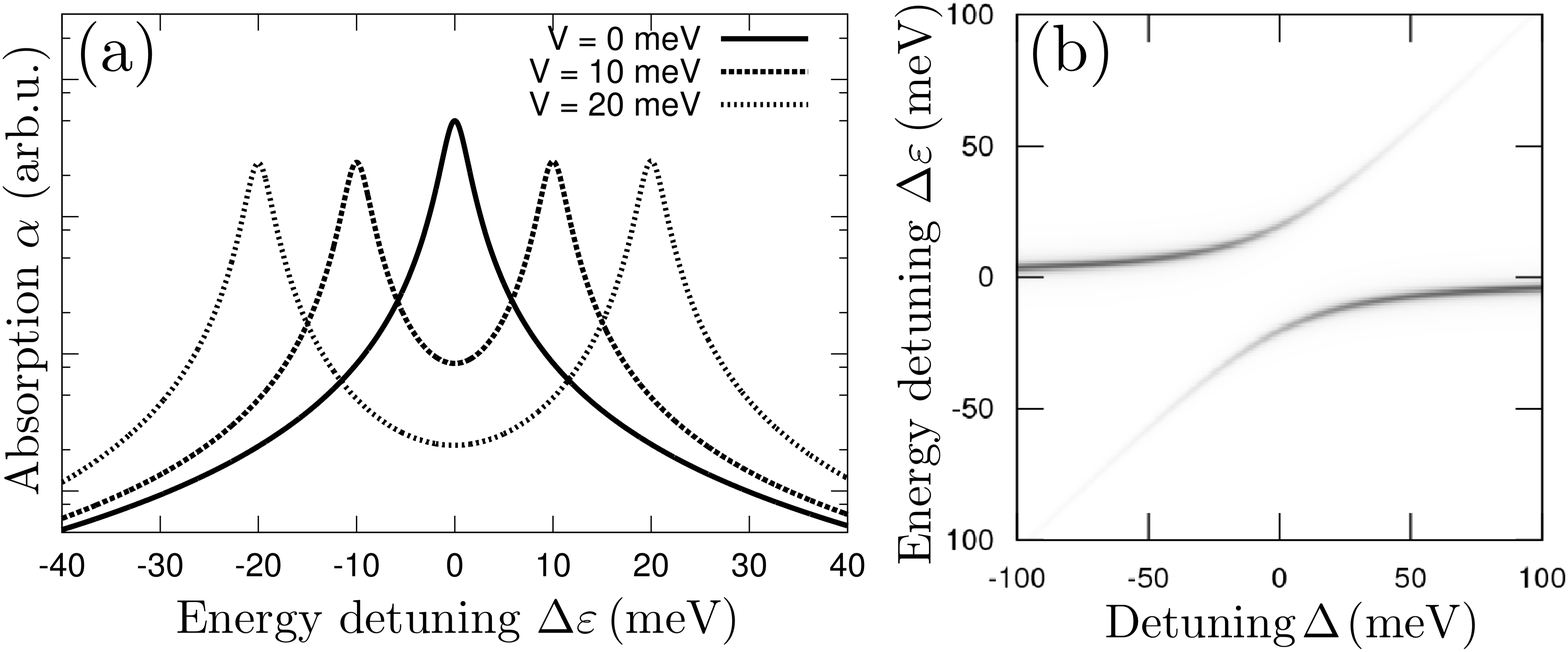}
\caption{(a) Linear absorption around the energy of the transition \state{v}{2}{\uparrow} $\rightarrow$
\state{c}{2}{\uparrow} for
$\Delta = 0$ [energy-conserving impact ionization, cf. Fig.~\ref{modelsystem}(c)] and different Coulomb coupling strengths
$V$. (b) Linear absorption for $V = 20~\text{meV}$ and varying detuning $\Delta$ and energy detuning $\Delta \varepsilon$, showing
an anticrossing between the single-exciton \dstate{c}{2}{\ua}{v}{2}{\ua}
and two-exciton state $|c 1 \!\! \ua c 1 \!\! \da v 1 \!\! \da v 2 \!\!
\ua\rangle$.}
\label{ANTI}
\end{figure}

The linear absorption spectrum of the system for different coupling strengths $V$ is shown in Fig.~\ref{ANTI}(a).
The double peak structure is the consequence of the anticrossing as illustrated in Fig.~\ref{ANTI}(b) for
different detunings $\Delta$.
The states are degenerate for vanishing Coulomb coupling; for non-vanishing coupling, new eigenstates which are
superpositions of the unperturbed states appear. In the following discussion, the system will be resonantly excited on the
higher energetic branch [cf. Fig.~\ref{ANTI}(b)] except where
explicitly noted otherwise. This excitation energy is found to be
$\hbar \omega \approx (\epsilon_{2\uparrow}^c -
\epsilon_{2\uparrow}^v + V)$ for $V\gtrsim3$~meV in the case of a vanishing detuning $\Delta$, which describes
energy-conserving impact ionization as considered here, cf. Fig.~\ref{modelsystem}(c).

\subsection{\label{structure}Structure of the dynamics}

Since the process of impact ionization is the focus of this study, we start the discussion of the
dynamical structure with
the expectation value $\langle p_{\text{II}} \rangle = \langle \edag{1\uparrow} \edag{1\downarrow} \h{1\downarrow}
\e{2\uparrow} \rangle$ [cf.
table \ref{operator_abbreviations}], which is the probability amplitude of the impact ionization process (generation of
two electrons in the conduction band via one electron in the conduction and one in the valence band):
\begin{align}
d_t  \langle p_{\text{II}} \rangle = &(i \Delta\omega^{(\text{II})} - \gamma_{\text{PD}})\langle p_{\text{II}} \rangle -
\frac{i}{\hbar}
M(t) \langle p_{\text{DE}} \rangle_ {-1} \nonumber \\
&+ \frac{i}{\hbar} V ( \langle f_{2 \uparrow}^{c} f^v_{1\downarrow}\rangle - \langle f^c_{1\uparrow} f^c_{1\downarrow} \rangle )
\label{EOM_II/AR}
\end{align}
with $\Delta \omega^{(\text{II})} = \hbar^{-1}(\epsilon_{1\uparrow}^c + \epsilon_{1\downarrow}^c  - \epsilon_{1\downarrow}^v -
\epsilon_{2\uparrow}^c)$. Here, the product of the dipole coupling element and the slowly varying envelope of the
electric field is denoted with $M(t) = \tilde{E}(t) M^{cv}_{2\uparrow 2\uparrow}$.
In the following, we write the classical light field $E(t)$ as a sum of its
positive ($e^{i\omega_L t}$) and negative ($e^{-i\omega_L t}$) frequency parts and abbreviate the
product of an expectation value $\langle A\rangle$ and a rotation of n times the laser frequency $e^{\pm i n \omega_L t}$
with
\begin{equation}
\langle A \rangle_{\pm n} = \langle A \rangle e^{\pm i n \omega_L t}\nonumber.
\end{equation}
We then apply a rotating wave approximation by neglecting fast rotating terms for each dynamical equation. For example,
in Eq.~(\ref{EOM_II/AR}), the term $\frac{i}{\hbar} M(t) \langle p_{\text{DE}} \rangle_ {+1}$ does not appear because
it oscillates approximately with the frequency $2 \omega_L$, whereas the other terms $\frac{i}{\hbar} M(t) \langle p_{\text{DE}}
\rangle_ {-1}$,
$\langle f_{2 \uparrow}^{c} f^v_{1\downarrow}\rangle$, and $\langle f^c_{1\uparrow} f^c_{1\downarrow} \rangle$ have no free
oscillations.

The impact ionization $\langle p_{\rm II} \rangle$ in Eq.~(\ref{EOM_II/AR}) is driven (i) via the optical field $M(t)$ by the double excitation amplitude
$\langle p_{\text{DE}}
\rangle_{-1}$ and (ii) via the Coulomb coupling $V$ by the density-density correlation $\langle f_{2 \uparrow}^{c}
f^v_{1\downarrow}\rangle$ of two electrons in the states \state{c}{2}{\uparrow} and \state{v}{1}{\downarrow} [cf.
Fig.~\ref{modelsystem}(c)]. The correlation $\langle f_{2 \uparrow}^{c}
f^v_{1\downarrow}\rangle$ describes the simultaneous existence of electrons in these two levels which is essential for the
scattering process of impact ionization to occur. Because the final states of the electrons undergoing impact ionization are
\state{c}{1}{\uparrow} and \state{c}{1}{\downarrow}, a simultaneous occupation of these two states $\langle f^c_{1\uparrow}
f^c_{1\downarrow} \rangle$ blocks the impact ionization process.
The source term due to the double
excitation process $\langle p_{\text{DE}} \rangle_ {-1}$, depicted in Fig.~\ref{modelsystem}(e), and mediated by the
coupling $M(t)$, depends on the presence of the light field and therefore changes $\langle p_{\text{II}} \rangle$ only
while the system
is interacting with the light pulse, in contrast to the Coulomb coupling $V$ which acts at all times. This fact will become
important in the analysis of the
occuring system dynamics.

The dynamical equations for the double excitation $\langle p_{\text{DE}} \rangle_ {-1}$ and the density-density correlations read
\begin{align}
d_t \langle p_{\text{DE}} \rangle_{-1} = &(i \Delta \omega^{(\text{DE})} - \gamma_{\text{PD}}) \langle p_{\text{DE}} \rangle_{-1} \label{EOM_PD} \\
&- \frac{i}{\hbar} M(t) \langle p_{\text{II}}\rangle + \frac{i}{\hbar} V \langle p_{2\uparrow} f^v_{1\downarrow} \rangle_{-1},
\nonumber
\end{align}
\begin{align}
d_t \langle f_{2 \uparrow}^{c} f^v_{1\downarrow} \rangle = &\frac{2}{\hbar} M(t) \Im \big( \langle p_{2\uparrow} f^v_{1\downarrow}
\rangle_{-1}\big) - \frac{2}{\hbar} V \Im \big( \langle p_{\text{II}} \rangle \big),
\label{EOM_fe2up_fv1do}
\end{align}
\begin{align}
d_t \langle f^c_{1\uparrow} f^c_{1\downarrow} \rangle = &\frac{2}{\hbar} V \Im \big( \langle p_{\text{II}} \rangle \big),
\label{EOM_fe1up_fe1do}
\end{align}
with $\Delta \omega^{(\text{DE})} = \hbar^{-1}(\epsilon_{1\uparrow}^c + \epsilon_{1\downarrow}^c  -
\epsilon_{1\downarrow}^v - \epsilon_{2\uparrow}^v - \hbar \omega_L)$.
The double excitation $\langle p_{\text{DE}} \rangle_{-1}$ is driven by the impact ionization $\langle p_{\text{II}}
\rangle$ via the light field $M(t)$ and by the term $\langle p_{2\uparrow} f^v_{1\downarrow} \rangle_{-1}$,
which describes the excitation of an electron from \state{v}{2}{\uparrow} to \state{c}{2}{\uparrow} correlated with an electron
density in the state \state{v}{1}{\downarrow}. The density-density correlation $\langle f_{2 \uparrow}^{c} f^v_{1\downarrow} \rangle$
is driven by this polarization $\langle
p_{2\uparrow} f^v_{1\downarrow} \rangle_{-1}$ and the impact ionization $\langle p_{\text{II}} \rangle$, while $\langle
f^c_{1\uparrow} f^c_{1\downarrow} \rangle$ is solely determined by impact ionization $\langle p_{\text{II}} \rangle$.

The system of equations is closed with the following two dynamical equations:
\begin{align}
d_t \langle p_{2\uparrow} f^v_{1\downarrow} \rangle_{-1} = &(i \Delta \omega^{(2\uparrow)} - \gamma_{\text{PD}})\langle p_{2\uparrow}
f^v_{1\downarrow} \rangle_{-1} + \frac{i}{\hbar} V \langle p_{\text{DE}} \rangle_{-1} \nonumber \\
&- \frac{i}{\hbar} M(t) ( \langle f_{2 \uparrow}^{c} f^v_{1\downarrow} \rangle - \langle f_{2 \uparrow}^{v} f^v_{1\downarrow}
\rangle), \label{p2up_fv1down}
\end{align}
\begin{align}
d_t \langle f_{2 \uparrow}^{v} f^v_{1\downarrow} \rangle = &-\frac{2}{\hbar}
M(t) \Im \big( \langle p_{2\uparrow} f^v_{1\downarrow} \rangle_{-1}\big) \label{EOM_fv2up}
\end{align}
with $\Delta \omega^{(2\uparrow)} = \hbar^{-1}(\epsilon_{2\uparrow}^c - \epsilon_{2\uparrow}^v - \hbar \omega_L)$.

\begin{figure}[htb]
\includegraphics[width=85mm,clip]{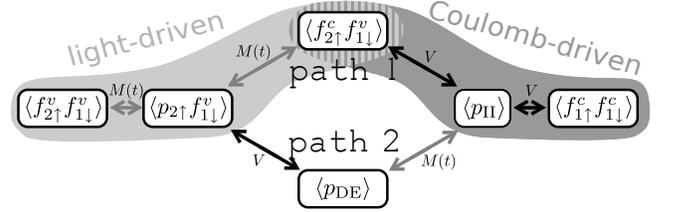}
\caption{Coupling scheme of the system dynamics given by Eqs.~(\ref{EOM_II/AR})-(\ref{EOM_fv2up}). Grey arrows depict
electron-light coupling and
black arrows indicate Coulomb coupling.}
\label{coupling}
\end{figure}

A better understanding of the dynamics can be obtained with the help of a graphical presentation of the
coupled expectation values, shown in Fig.~\ref{coupling}. We start
  from the initial condition that only the density
correlations in the valence band are equal to one, i.e. $\langle
f^v_{2\uparrow} f^v_{1\downarrow} \rangle = 1$, while all other density
correlations and polarizations vanish.
The dynamics can essentially be divided into two paths, leading from the optically excited polarization
$\langle
p_{2\uparrow} f^v_{1\downarrow} \rangle_{-1}$  to the impact ionization amplitude $\langle p_{\text{II}} \rangle$, which is
directly related to the two-exciton state $\langle f^c_{1\uparrow} f^c_{1\downarrow}
\rangle$: one via
the density-density correlation $\langle f_{2 \uparrow}^{c} f^v_{1\downarrow} \rangle$ (\textit{path 1}) and one via the
correlation describing the double excitation $\langle p_{\text{DE}} \rangle$ (\textit{path 2}). \textit{Path 1} can be decomposed
into two contributions, each coupling two density-density
correlations via a corresponding polarization ($\langle p_{2\uparrow}
f^v_{1\downarrow} \rangle_{-1}$, $\langle \pII \rangle$).
The polarization itself is driven by an inversion between the corresponding
density-density correlations
($\langle f_{2 \uparrow}^{c} f^v_{1\downarrow} \rangle - \langle f_{2 \uparrow}^{v} f^v_{1\downarrow}
\rangle$,$\langle f_{2 \uparrow}^{c} f^v_{1\downarrow}\rangle - \langle f^c_{1\uparrow} f^c_{1\downarrow} \rangle$). If only the
light-matter-coupling $M(t)$ is considered, the contribution is comprised of the polarization $\langle p_{2\uparrow}
f^v_{1\downarrow}
\rangle_{-1}$ and the density-density correlations $\langle f_{2 \uparrow}^{v} f^v_{1\downarrow} \rangle$ and $\langle f_{2
\uparrow}^{c} f^v_{1\downarrow}
\rangle$. Including the Coulomb coupling $V$, the contribution consists of the impact ionization $\langle p_{\text{II}}
\rangle$ and the
density-density correlations $\langle f_{2 \uparrow}^{c} f^v_{1\downarrow} \rangle$ and $\langle f_{1 \uparrow}^{c} f^c_{1\downarrow}
\rangle$. \textit{Path 2} links the polarizations $\langle p_{2\uparrow} f^v_{1\downarrow} \rangle_{-1}$ and $\langle p_{\text{II}}
\rangle$ via the double excitation $\langle p_{\text{DE}} \rangle_{-1}$.

If the typical time scales $\hbar/V$ and $\tau$ (pulse duration) of the two interactions $V$ and $M(t)$
obey the relation $\tau \lesssim \hbar/V$, the dynamics can be divided into sequential interaction
steps: For example, for a short light-matter ($\tau$) and a comparably long Coulomb interaction time
$(\hbar/V)$, 
the dynamics is first restricted solely to the left
contribution of path 1. For longer times, the right contribution of path 1 is activated by the non-vanishing
density-density
correlation $\langle f^c_{2\uparrow} f^v_{1\downarrow}\rangle$, while path 2 remains dormant due to
vanishing
intermittent polarizations. For a light-matter interaction strongly exceeding the Coulomb interaction time scale, i.e.,
$\tau \gg \hbar/V$, the
dynamics is comprised of an
interference of both paths.

A case where path 2 dominates the dynamics while path 1 remains dormant is not possible, as
can be seen in Fig.~\ref{coupling}. Let us assume that $\hbar/V$ is negligible compared to $\tau$. Even though $\langle \pDE \rangle$
of path 2 is then driven on a much shorter time scale than the density-density correlation $\langle
f^c_{2\uparrow} f^v_{1\downarrow} \rangle$ of path 1 via the polarization $\langle p_{2\uparrow} f^v_{1\downarrow} \rangle$, this
process is compensated by the slower time scale of the second step between $\langle \pDE \rangle$ and $\langle \pII \rangle$ as
compared to the case for the density-density correlation $\langle f^c_{2\uparrow} f^v_{1\downarrow} \rangle$. Thus, no qualitative
difference in the behavior is expected when taking the limiting case of $\tau \gg \hbar/V$, which is just due to
the fact that the Coulomb interaction acts at all times.

We will come back to the dynamical structure via the two paths in the following sections to explain the numerical
results.

\section{\label{carrMult}Carrier Multiplication}

We now turn to the central focus of the article, the investigation of the ultrafast dynamics of carrier multiplication. As a measure
of the process, we define the carrier multiplication $\text{CM}$ as the ratio of the total carrier occupation in the conduction band
$\langle
f_c\rangle$ and its light-induced part $\langle f_c \rangle\big|_{\text{light}}$, i.e., the conduction band carrier
density which is excited solely by the light field:
\begin{equation}
\text{CM} = \frac{\langle f_c\rangle}{\langle f_c \rangle\big|_{\text{light}}} \text{ .}
\end{equation}
The quantity $\text{CM}$ is equal to one if no {\em additional}
carriers are created by the Coulomb
coupling ($\langle f_c \rangle = \langle f_c \rangle\big|_{\text{light}}$). Its maximal value in the model system
treated in this paper (cf. Sec.~\ref{model}) is $2$. This results from the fact that the
electronic excess energy is equal to two times the band gap energy.
We will now derive the explicit expressions for the total carrier occupation in the conduction band $\langle f_c \rangle$, its
Coulomb-induced part $\langle f_c \rangle\big|_{\text{Coul}}$ and its light-induced part $\langle f_c \rangle\big|_{\text{light}}$.

The total carrier occupation in the conduction band reads
\begin{equation}
\langle f_c \rangle = \sum_{\lambda,s} \langle a_{c \lambda s}^{\dag} a_{c \lambda s}^{} \rangle = \langle
f_{2 \uparrow}^{c} \rangle + \langle f_{1\uparrow}^{c} \rangle + \langle f_{1 \downarrow}^{c} \rangle \hspace{2mm}
\text{.}
\label{def_total}
\end{equation}
The occupation $\langle f_{2\downarrow}^c \rangle$ is neither driven by the external optical field nor by
Coulomb scattering and
therefore remains zero.
Considering the dynamical equation for $\langle f_c \rangle$,
\begin{equation}
d_t \langle f_c \rangle = \underbrace{\frac{2}{\hbar} M(t) \Im \big( \langle p_{2 \uparrow} \rangle_{-1}
\big)}_{d_t \langle f_c \rangle\big|_{\text{light}}}
+ \underbrace{\frac{2}{\hbar} V \Im \big( \langle
p_{\text{II}} \rangle \big)}_{d_t \langle f_c \rangle\big|_{\text{Coul}}} \hspace{2mm}\text{,}
\label{eq_CM}
\end{equation}
two source terms $\langle f_c \rangle\big|_{\text{light}}$ and $\langle f_c \rangle\big|_{\text{Coul}}$ can be identified,
relating to the light-matter coupling $M(t)$ and the impact
  ionization transition probability, respectively. Thus, the
  carrier multiplication $\text{CM}$ can be computed.

In the following, we analyze carrier multiplication for both the energy-conserving and the non-energy-conserving case
numerically, using the parameters in table~\ref{table_parameter} if not noted otherwise.
\begin{table}[htb]
\begin{tabular}{c c c}
\hline
\hline
Parameter &\hspace*{1cm} &Value \\
\hline
\hline
$\epsilon^v_{2\uparrow},\epsilon^v_{2\downarrow}$ && $-400 \text{ meV}$ \\
$\epsilon^v_{1\uparrow},\epsilon^v_{1\downarrow}$ && $0 \text{ meV}$ \\
$\epsilon^c_{1\uparrow},\epsilon^c_{1\downarrow}$ && $400 \text{ meV}$ \\
$\epsilon^c_{2\uparrow},\epsilon^c_{2\downarrow}$ && $800 \text{ meV}$ \\
$V$ && $20 \text{ meV}$ \\
$\gamma_{\text{PD}}$ && $1/(500 \text{ fs})$\\
\hline
\hline
\label{parameters}
\end{tabular}
\caption{The electronic energy structure and the Coulomb coupling elements are taken from Ref.~\onlinecite{Shabaev2006} for a PbSe quantum dot of diameter $d$ = 5 nm. This system provides a small band gap and allows energy conserving impact ionization and Auger recombination processes.}
\label{table_parameter}
\end{table}

\subsection{\label{genProps}General Properties of our Model}

We first discuss the general relation between the total carrier occupation and its light- and Coulomb-induced parts in our model 
system. This will be useful
in understanding the dynamical behavior of these quantities in the next section.

The inversion driving the impact ionization [cf. Eq.~\ref{EOM_II/AR}] and Auger recombination can be expressed in terms of the light- and
Coulomb-induced parts of the total carrier occupation:
\begin{align}
\langle f^c_{2\uparrow} f^v_{1\downarrow} \rangle - \langle f^c_{1\uparrow} f^c_{1\downarrow} \rangle =
\langle f_c \rangle\big|_{\text{light}} - 2 \langle f_c \rangle\big|_{\text{Coul}}\text{ .}
\label{INV_Coul}
\end{align}
Here, we employed the dynamical equations (\ref{EOM_fe2up_fv1do},\ref{EOM_fe1up_fe1do}) and the
definitions of $\langle f_c \rangle\big|_{\text{light}}$ and $\langle f_c \rangle\big|_{\text{Coul}}$ as well as their initial values.
For a vanishing inversion, the ratio of $\langle f_c \rangle\big|_{\text{Coul}}$ and $\langle f_c \rangle\big|_{\text{light}}$ 
is obtained as
\begin{align}
\langle f_c \rangle\big|_{\text{Coul}} = 1/2 \langle f_c \rangle\big|_{\text{light}}\text{ .}
\end{align}
This is equivalent to a carrier multiplication $\text{CM} = 1.5$, which is reached in a quasi-stationary limit, where all
polarizations (in particular, the impact ionization and Auger recombination amplitudes) vanish. 

In our model system, 
this value is obtained on long time scales independent of the exciting pulse parameters, as will be seen in the next sections. This is due to the 
fact that we do not include relaxation effects such as phonon-assisted intraband relaxation, since we want to focus on the ultrafast
time dynamics of 
carrier multiplication. Furthermore, we will see that, in contrast to first expectations, reaching this quasi-stationary state of
the system does {\em not} solely depend on the damping of the participating polarizations by
pure-dephasing processes, but can be reached on an ultrafast timescale by control of the pulse duration $\tau$.

A complementary analysis was performed in Ref.~\onlinecite{Shabaev2006} for the absorption bleaching, focusing on the
relation
between the Coulomb interaction strength and the relaxation rates of the single-exciton and
two-exciton states. Here, we focus on the relation of the Coulomb coupling strength to the
externally controllable pulse length, allowing an external control in the detection of carrier multiplication.
Since we do not consider relaxation processes, we are thus in the regime where oscillations are observed in the absorption
bleaching in Ref.~\onlinecite{Shabaev2006}.

\subsection{\label{energy_conserving}Energy-Conserving Case}

We begin our investigation of the dynamics by considering energy-conserving impact ionization, i.e., we assume the
energetic structure given in table~\ref{table_parameter}, where $\epsilon^c_{2\uparrow} - \epsilon^c_{1\uparrow} =
\epsilon^c_{1\downarrow} - \epsilon^v_{1\downarrow}$. We differentiate between two different regimes: (i) the light-matter coupling
occurs on a shorter time scale than the Coulomb coupling, i.e., the pulse duration $\tau$ is shorter than the
typical Coulomb interaction time $(\hbar/V)$, $\tau \lesssim \hbar/V$ and (ii) the light-matter coupling strongly exceeds
the Coulomb coupling, $\tau \gg \hbar/V$.

\begin{figure}[htb]
\includegraphics[width=60mm,clip]{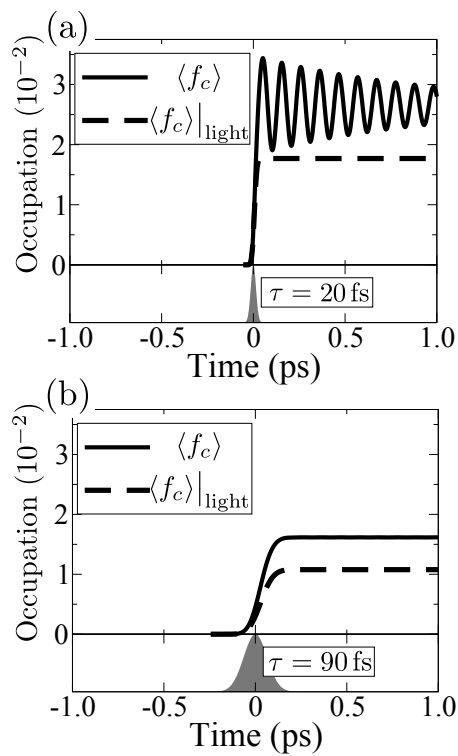}
\caption{Dynamics of the conduction band carrier occupation $\langle f_{c} \rangle$ and the contribution due to the
light field $\langle f_{c} \rangle\big|_{\text{light}}$ for a
pulsed excitation with a pulse area of $0.1\pi$ and pulse
length of (a) $\tau = 20 \text{ fs}$ and (b) $\tau = 90 \text{ fs}$.}
\label{cmdynamics1}
\end{figure}

(i) \textit{$\tau \lesssim \hbar/V$}: Figure~\ref{cmdynamics1}(a) shows the conduction band carrier occupation $\langle f_c \rangle$
and the contribution from the
light-matter interaction $\langle f_c \rangle\big|_{\text{light}}$ for an excitation with a $20$ fs pulse with a pulse area of $0.1
\pi$.
The light pulse induces a density in the conduction band $\langle f_c \rangle\big|_{\text{light}}$ at
$t\approx 0$~ps. After the pulse, $\langle f_c \rangle\big|_{\text{light}}$ remains constant (dashed line) and
is exceeded by the total carrier density (solid line) $\langle f_c \rangle$, which indicates carrier multiplication.
$\langle f_c \rangle$ exhibits clear oscillations above the level of $\langle f_c \rangle\big|_{\text{light}}$.

As discussed in Sec.~\ref{structure}, the build-up of the light- and Coulomb-induced carrier density occurs {\em
sequentially} via path 1 depicted in Fig.~\ref{coupling}. For the considered parameter range, on a short time
scale, determined by the light-matter interaction $\tau$, the light field induces the density-density correlation $\langle
f^c_{2\uparrow} f^v_{1\downarrow} \rangle$ which, on a time scale determined by the Coulomb coupling $\hbar/V$, creates the
density-density correlation $\langle f^c_{1\uparrow} f^c_{1\downarrow} \rangle$ via impact ionization. During this build-up, impact
ionization becomes increasingly inhibited because the inversion
between the density-density correlations $\langle f^c_{1\uparrow} f^c_{1\downarrow} \rangle$ and $\langle f^c_{2\uparrow}
f^v_{1\downarrow} \rangle$ tends to zero, leading to a slowdown of impact ionization until a first maximum of $\langle f_c \rangle$
is reached, i.e., a state which favors Auger recombination over impact ionization.
This leads to a subsequent decrease of the total carrier occupation towards a local temporal minimum until impact ionization starts
to dominate again.
The damping of the resulting oscillation of the total carrier occupation $\langle f_c \rangle$ is determined solely
by the pure dephasing of the system.

The basic physical picture of the oscillation is the anticrossing described in Sec.~\ref{structure}:
The spectrally broad pulse simultaneously excites both eigenstates, creating an inversion between the unperturbed states, i.e., 
between the correlations 
$\langle f_{2 \uparrow}^{c} f^v_{1\downarrow}\rangle$ and $\langle f^c_{1\uparrow} f^c_{1\downarrow} \rangle$ [cf. Eq.~(\ref{EOM_II/AR})].
The limit of $\text{CM} = 1.5$ is reached as soon as the pure dephasing has damped the participating polarizations.

(ii) \textit{$\tau \gg \hbar/V$}: Under excitation with a longer light pulse ($\tau$ = 90 fs), the time scale of the light-matter 
coupling strongly exceeds the Coulomb interaction ($\tau = 90 \text{ fs}$, $\hbar/V \approx 35 \text{
fs}$). This prevents carrier multiplication, i.e., the
creation of the density-density correlation $\langle f^c_{1\uparrow} f^c_{1\downarrow} \rangle$), to occur as a sequential process of initial
light excitation and following impact ionization alone. Now, the paths
depicted in Fig.~\ref{coupling} describing a sequential (path 1) and a simultaneous (path 2) creation of the
density-density correlation $\langle f^c_{1\uparrow} f^c_{1\downarrow} \rangle$ contribute both and lead to a concurrent
growth of the light-induced
$\langle f_c \rangle\big|_{\text{light}}$ and Coulomb-induced $\langle f_c \rangle\big|_{\text{Coul}}$ carrier densities in the
conduction band. Figure~\ref{cmdynamics1}(b) shows that carrier multiplication still occurs, even though no oscillations
(no subsequent impact ionization and Auger recombination)
are found in the total carrier density $\langle f_c \rangle$. This is due to a balance between the densities driving
the impact ionization, and therefore a vanishing inversion, resulting from the interplay of paths 1 and 2.
This balance causes the carrier multiplication limit of $\text{CM} = 1.5$ to be reached nearly simultaneously with the exciting
light pulse on a much shorter timescale than in the case of short pulse excitation.

\subsection{Non-energy-conserving Case}

\begin{figure}[htb]
\includegraphics[width=85mm,clip]{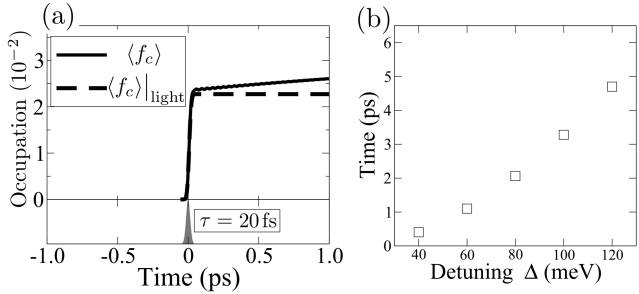}
\caption{(a) Time evolution of the total carrier occupation in the conduction band $\langle f_{c} \rangle$ and its
light-induced part
$\langle f_{c} \rangle\big|_{\text{light}}$ for a detuning of $\Delta = 100$ meV between the Coulomb-coupled states and an excitation with a 20~fs, 0.1$\pi$
pulse. (b)
Impact ionization times, defined as the time of the system to reach a carrier multiplication factor of
$\text{CM} = 1.316 \, (= 1+0.5(1-e^{-1}))$ over the detuning
$|\Delta|$ for an excitation resonant on the single-exciton energy $\hbar \omega_L = (\epsilon^c_{2\uparrow} -
\epsilon^v_{2\uparrow})$.
}
\label{DETUNINGS}
\end{figure}

When the Coulomb-coupled single-exciton and two-exciton states are off-resonant, i.e., $\epsilon^c_{2\uparrow} -
\epsilon^c_{1\uparrow} \neq \epsilon^c_{1\downarrow} - \epsilon^v_{1\downarrow}$, the influence of impact ionization
is strongly suppressed, because energy conservation is violated by this scattering process. To investigate this
quantitatively for our Coulomb parameters (cf. table~\ref{table_parameter}), the time evolution of the conduction band densities
$\langle f_c \rangle$ and  $\langle f_c \rangle\big|_{\text{light}}$ is shown in Fig.~\ref{DETUNINGS}(a) for an exemplary
detuning $\Delta = 100$ meV between the Coulomb-coupled states and a pulse length of $20 \text{ fs}$.
The detuning $\Delta$ is achieved by varying the energy $\epsilon^v_{1\downarrow}$, while the other parameters remain fixed.
 The system is excited resonantly on the
single-exciton energy $\epsilon^c_{2 \uparrow} - \epsilon^v_{2 \uparrow}$.

The Coulomb-induced conduction band density $\langle f_c \rangle\big|_{\text{Coul}}$, and thus the process of carrier multiplication, is
negligibly weak on an ultrafast subpicosecond time scale. For very long time scales, the $\text{CM}$ again
approaches 1.5, as discussed in Sec.~\ref{genProps}, assuming that no radiative recombination and phonon-assisted intraband
relaxation are considered.
However, when these processes are considered, which act on a picosecond time scale in the case of
phonon-assisted relaxation, the carrier multiplication will dramatically decrease below $\text{CM} = 1.5$.

To investigate the influence of off-resonant Coulomb scattering processes on the timescale of carrier multiplication, we plot the time 
in which a value $\text{CM} = 1.316$, corresponding to the exponential growth value of the $\text{CM}$ maximum, is reached
over the detuning $\Delta$ in Fig.~\ref{DETUNINGS}(b).
Carrier multiplication times of a few picoseconds, which lie in the range of typical phonon-assisted relaxation times,\cite{Nozik2010} 
occur for a detuning $\Delta \gtrsim 100 \text{meV}$.

Thus, for sufficient detuning between the Coulomb-coupled states, impact
ionization is expected to be negligible. This result can also be used to justify neglecting impact ionization scattering in the valence band in 
material systems 
which show a large difference in the  effective masses in the conduction and valence band.
Nevertheless, it should be noted that this result is crucially dependent on the Coulomb coupling strength $V$. A stronger
coupling would raise the importance of non-energy-conserving impact ionization for a given detuning~$\Delta$.

\subsection{Inclusion of valence band impact ionization}

To include the influence of impact ionization and Auger recombination in the valence band [Fig.~\ref{modelsystem}(d)] in our dynamics,
we generalize the Coulomb Hamiltonian $H_{\text{Coul}}$:
\begin{align}
H_{\text{Coul}}=
&V^{\text{II}}
\hspace{1.2mm} \edag{1 \uparrow} \edag{1 \downarrow} \h{1 \downarrow} \e{2 \uparrow}
+ V^{\text{AR}}
\hspace{1.2mm} \edag{2 \uparrow} \hdag{1 \downarrow} \e{1 \downarrow} \e{1 \uparrow}
\nonumber
\\
+&\bar{V}^{\text{II}}
\hspace{1.2mm} \hdag{2 \uparrow} \edag{1 \downarrow} \h{1 \downarrow} \h{1 \uparrow}
+ \bar{V}^{\text{AR}}
\hspace{1.2mm} \hdag{1 \uparrow} \hdag{1 \downarrow} \e{1 \downarrow} \h{2 \uparrow}\text{ ,}
\nonumber
\end{align}
with the coupling elements $\bar{V}^{\text{II}} = V_{2 \uparrow 1 \downarrow 1 \downarrow 1 \uparrow}^{v \hspace{1.4mm} c
\hspace{1.4mm} v \hspace{1.4mm} v} = (\bar{V}^{\text{AR}})^* $.
This introduces additional valence band scattering correlations as well as correlations linking both bands.

\begin{figure}[htb]
\includegraphics[width=60mm,clip]{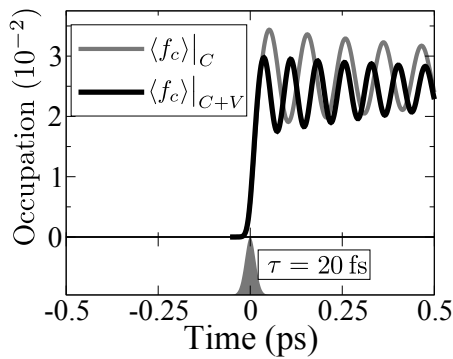}
\caption{
Time evolution of the total carrier occupation in the conduction band with ($\langle f_{c} \rangle\big|_{C+V}$) and without 
($\langle f_{c} \rangle\big|_{C}$) impact ionization processes in the valence band. The system is again excited on the higher
energetic branch. Differences between the oscillation frequencies and the mean values of the total occupation $\langle f_c \rangle$
result from a changed effective Coulomb coupling constant.}
\label{FIG_without_with}
\end{figure}

Figure~\ref{FIG_without_with} shows the corresponding dynamics with ($\langle f_{c} \rangle\big|_{C+V}$) and without 
($\langle f_{c} \rangle\big|_{C}$) impact ionization of valence band
electrons for a short pulse
excitation. The system is again excited resonant on the higher
energetic branch, and the conduction and valence band impact ionization coupling elements are taken to be equal, $V =
\bar{V} = 20 \text{ meV}$.
The occuring oscillation describing impact ionization and Auger recombination 
has a higher frequency for the inclusion of both processes which corresponds to a larger effective Coulomb
coupling $V_{\text{eff}}(V,\bar{V})$. For the long pulse excitation, no significant differences are found.

\section{Time-resolved optical emission}

In a next step, we investigate the obtained dynamical results in 
the time-resolved optical emission of the quantum dot during the impact ionization process, opening an alternative approach to
measure MEG. In order to describe this quantum-optical signal, we consider the fully quantized light-matter interaction Hamiltonian
\begin{align}
H_{\text{QO}} = \sum_{\textbf k} M^{\textbf k}_{2\uparrow} \edag{2\uparrow} \h{2\uparrow} c_{\textbf k}^{\phantom \dag} + \text{h.a.},
\end{align}
with the electron-photon coupling element $M^{\textbf k}_{2\uparrow} =  i \sqrt{\frac{\hbar \omega_{k}}{2 \epsilon_0 V}} M_{2\uparrow
2\uparrow}^{cv}$ in
dipole approximation; $c_{\textbf k}^{(\dag)}$ is the photonic annihilation (creation) operator in mode ${\textbf k}$.
Analogous to the consideration of the semiclassical light-matter interaction, a rotating-wave
approximation is applied on the frequency of the external excitation. The hierarchy of dynamical equations stemming from
the
electron-photon interaction is truncated on the two-photon level, i.e., two-photon assisted electronic correlations are neglected.

The time-resolved emission spectrum at an observation frequency $\omega_{\textbf k_s}$ at a distance $z$ to the detector is 
given by\cite{Kabuss2010}
\begin{align}
\begin{split}
S(z\vec{e}_z,\omega_{k_s},t) = \sum_{{\textbf k_1},{\textbf k_2}} \hbar \frac{\sqrt{\omega_{k_1} \omega_{k_2}}}{2 \epsilon_0 c^2
\Omega}
\langle c_{{\textbf k_1}}^\dag c_{{\textbf k_2}}\rangle e^{-i(k_1 - k_2)z}
\\
\times
e^{-\left[(k_1 - k_s)c\Delta t/\sqrt{2}\right]^2}e^{-\left[(k_2 - k_s)c\Delta t/\sqrt{2}\right]^2}\hspace{2mm},
\end{split}
\label{TIME-SIG}
\end{align}
with the time uncertainty of the detector $\Delta t$, the quantization volume $\Omega$, the
speed of light $c$, and the vacuum dielectric constant $\epsilon_0$. The angular frequencies $\omega_i$ and the momenta $k_i$ are
related via the linear dispersion relation $\omega_i = c k_i$. The time uncertainty $\Delta t$ has to be equal to or smaller
than the pulse length $\tau$ in order to resolve the time evolution of the exciting light pulse.

\begin{figure}[htb]
\includegraphics[width=85mm,clip]{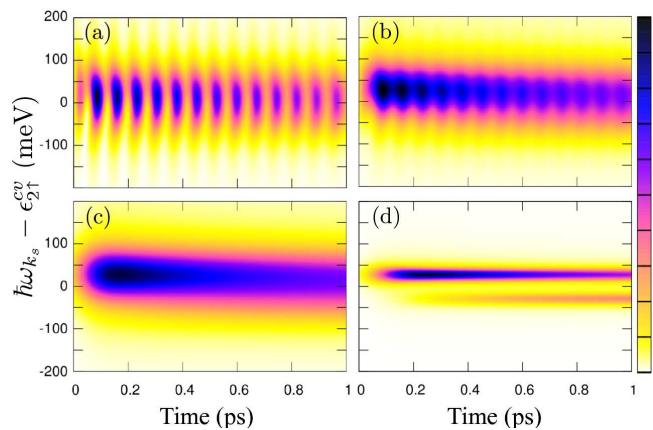}
\caption{(Color online) Time-resolved optical emission spectra around the energy of the single-exciton $\epsilon_{2\uparrow}^{cv} =
(\epsilon_{2\uparrow}^c - \epsilon_{2\uparrow}^v)$ for an excitation with a pulse area of $0.1 \pi$ and pulse length (a) $\tau =
20\text{ fs}$, (b) $\tau = 40\text{ fs}$, and (c),(d) $\tau = 90\text{ fs}$. (a)-(c) The time uncertainty of the detector is set to
$\Delta t = 20\text{ fs}$. A transition from oscillating to non-oscillating behavior occurs, corresponding to the carrier dynamics
discussed earlier. (d) For a larger time uncertainty $\Delta t = 20\text{ fs}$, the better energetic resolution reveals the
substructure of the fluorescence corresponding to the observed anticrossing in the linear absorption.
}
\label{FIG_QO}
\end{figure}

In Figs.~\ref{FIG_QO}(a-c), the resulting time-resolved emission spectrum is shown for increasing pulse lengths $\tau$ = 20 fs, 40 fs,
and 90 fs. The time uncertainty $\Delta t$ is set to 20 fs in order to resolve the temporal dynamics. In the calculations, the full
system including impact ionization and Auger recombination in both the conduction and valence band is considered.
For a short pulse excitation [Fig.~\ref{FIG_QO}(a)], oscillations in the emission are found, corresponding to the impact
ionization-induced carrier dynamics in Fig.~\ref{cmdynamics1}(a). These oscillations become less pronounced for an intermediate
pulse length
[Fig.~\ref{FIG_QO}(b)] and disappear for an excitation with a long pulse [Fig.~\ref{FIG_QO}(c)]. 
Thus, the quantum-kinetic signatures, resulting from the interplay between the light-matter and the Coulomb interaction, are
recovered directly in the time-resolved optical emission.

To recover the energetic structure of the QD, the time uncertainty of the detector is increased in Fig.~\ref{FIG_QO}(d) to allow for a sufficient energetic resolution. The observed energetic splitting of the fluorescence line corresponds to the anticrossing in the
linear absorption (cf.~Fig.~\ref{ANTI}). The earlier onset of the signal at the energetically higher branch results from the Rayleigh
signal of the exciting light pulse.

\section{Conclusion}

A quantum-kinetic approach to the dynamics of carrier multiplication in quantum dots was presented, focusing on the interplay between the light-matter
and the Coulomb interaction on ultrafast timescales, where intraband relaxation is of minor importance.
The analysis of the underlying dynamics revealed two
different paths leading to multiple carriers in the conduction band, controllable via the coupling parameters of the two interactions:
the duration of the exciting light pulse and the Coulomb interaction time.

We observe the occurence of oscillations for certain externally controllable excitation parameters. These oscillations result from the
interplay of impact ionization and Auger recombination and could in principle be used to detect carrier multiplication before
relaxation processes occur. The corresponding distinct signatures in the time-resolved optical emission spectrum open the
possibility for an alternative approach to measure multi-exciton generation in quantum dots. In particular,
they might offer the possibility to distinguish carrier multiplication in quantum dots by Auger-type processes from other processes, such as multiphoton excitation, uncontrolled photocharging, and influences of the surface chemistry, which can lead to ambiguities in the analysis of excitonic decay dynamics.\cite{Pijpers2008,McGuire2010,Tyagi2011}

It was also shown that the time dynamics does not qualitatively change when neglecting impact ionization scattering in the valence
band, resulting only in small modifications in the carrier occupations. Furthermore, the investigation of non-energy-conserving impact
ionization and Auger recombination highlights how crucial resonance is for the impact ionization process.

\vfill

\begin{acknowledgments}
We acknowledge financial support by the Deutsche Forschungsgemeinschaft via GRK1558 "Nonequilibrium Collective Dynamics in Condensed
Matter and Biological Systems".
\end{acknowledgments}

\appendix*
\section{Dynamical equations}

In the following, we present the dynamical equations used in the calculations. We restrict the equations to impact ionization and Auger recombination processes in the conduction band.

\subsection{Electronic system}

As in Sec.~\ref{structure}, we begin with the dynamical equation of the impact ionization amplitude:
\begin{align}
d_t \langle p_{\text{II}}  \rangle = &\frac{i}{\hbar} ( \epsilon^c_{1\ua} + \epsilon^c_{1\da} - \epsilon^v_{1\da} - \epsilon^c_{2\ua} + i \hbar \gamma_{\text{PD}} ) \langle p_{\text{II}} \rangle \nonumber \\
&-\frac{i}{\hbar} M(t) \langle p_{\text{DE}}  \rangle_{-1}
\nonumber
+ \frac{i}{\hbar} V \langle f^c_{2\uparrow} f^c_{1\uparrow} f^c_{1\downarrow} \rangle
\\
&+ \frac{i}{\hbar} V \big( \langle f^c_{2\uparrow} f^v_{1\downarrow}  \rangle
- \langle f^c_{1\uparrow} f^c_{1\downarrow}  \rangle \big) \nonumber \\ 
&- \frac{i}{\hbar} V \big( \langle f^c_{2\uparrow} f^c_{1\uparrow} f^v_{1\downarrow} \rangle + \langle f^c_{2\uparrow}
f^c_{1\downarrow} f^v_{1\downarrow} \rangle \big)
\end{align}
Next, we continue with the two-electron correlations:
\begin{align}
d_t \langle p_{\text{DE}}  \rangle_{-1}	= &\frac{i}{\hbar} ( \epsilon^c_{1\ua} + \epsilon^c_{1\da} - \epsilon^v_{1\da} -
\epsilon^v_{2\ua}  -\hbar \omega_{L} + i \hbar \gamma_{\text{PD}}) \langle p_{\text{DE}}  \rangle_{-1} \nonumber
\\
\nonumber
&- \frac{i}{\hbar} M(t) \langle p_{\text{II}}  \rangle + \frac{i}{\hbar} V \langle f^v_{1\downarrow} p_{2\uparrow} \rangle_{-1} \\
\nonumber
&+ \frac{i}{\hbar} V \big(\langle f^c_{1\uparrow} f^c_{1\downarrow} p_{2\uparrow} \rangle_{-1}
-
\langle f^c_{1\downarrow}f^v_{1\downarrow} p_{2\uparrow} \rangle_{-1}\big)
\\
&- \frac{i}{\hbar} V
\langle f^c_{1\uparrow} f^v_{1\downarrow} p_{2\uparrow} \rangle_{-1}
\end{align}
\begin{align}
d_t \langle f^c_{2\uparrow} f^v_{1\downarrow}  \rangle	= \hzwei M(t) \imag{\langle f^v_{1\downarrow} p_{2\uparrow}  \rangle_{-1}} - \hzwei V \imag{\langle p_{\text{II}} \rangle} 
\end{align}
\begin{align}
d_t \langle f^c_{1\uparrow} f^c_{1\downarrow}  \rangle	= \hzwei V \imag{\langle p_{\text{II}} \rangle}
\end{align}
\begin{align}
d_t \langle f^v_{1\downarrow} p_{2\uparrow}  \rangle_{-1} = &\frac{i}{\hbar} ( \epsilon^c_{2\ua} - \epsilon^v_{2\ua}  -\hbar
\omega_{L} + i \hbar \gamma_{\text{PD}}) \langle f^v_{1\downarrow}
p_{2\uparrow} \rangle_{-1} \nonumber
\\ 
&+ \frac{i}{\hbar}  V \big(\langle p_{\text{DE}}  \rangle_{-1} - \langle f^c_{2\uparrow} p_{\text{DE}} \rangle_{-1}\big) \nonumber \\
&+ \frac{i}{\hbar} M(t) \big(\langle f^v_{2\uparrow} f^v_{1\downarrow} 
\rangle - \langle f^c_{2\uparrow} f^v_{1\downarrow}  \rangle\big) 
\end{align}
\begin{align}
d_t \langle f^v_{2\uparrow} f^v_{1\downarrow}  \rangle	= - \hzwei M(t) \imag{\langle f^v_{1\downarrow} p_{2\uparrow} \rangle_{-1}} - \hzwei V \imag{\langle f^v_{2\uparrow} p_{\text{II}} \rangle}
\end{align}

The equations necessary to define the total carrier occupation and its parts are:
\begin{align}
d_t \langle p_{2\uparrow}  \rangle_{-1}	= &\frac{i}{\hbar} ( \epsilon^c_{2\ua} - \epsilon^v_{2\ua}  -\hbar \omega_{L} + i \hbar
\gamma_{\text{PD}}) \langle p_{2\uparrow}  \rangle_{-1} \\
&+ \frac{i}{\hbar} M(t) \big(\langle f^v_{2\uparrow}  \rangle - \langle f^c_{2\uparrow}  \rangle\big) + \frac{i}{\hbar}  V \langle
p_{\text{DE}}  \rangle_{-1} \nonumber
\end{align}
\begin{align}
d_t \langle f^v_{2\uparrow}  \rangle = - \hzwei M(t) \imag{ \langle p_{2\uparrow}  \rangle_{-1}} \label{App_EOM_fv2up}
\end{align}
\begin{align}
d_t \langle f^c_{2\uparrow}  \rangle	= \hzwei M(t) \imag{\langle p_{2\uparrow}  \rangle_{-1}} - \hzwei V \imag{\langle
p_{\text{II}}  \rangle}\label{App_EOM_fe2up}
\end{align}
\begin{align}
d_t \langle f^c_{1\uparrow} \rangle = \hzwei V \imag{\langle p_{\text{II}} \rangle}
\end{align}
\begin{align}
d_t \langle f^c_{1\downarrow} \rangle = \hzwei V \imag{\langle p_{\text{II}} \rangle}
\end{align}

\begin{align}
d_t&\langle f^c_{1\uparrow} f^c_{1\downarrow} p_{2\uparrow}  \rangle_{-1} = \frac{i}{\hbar}  V \langle f^c_{2\uparrow} p_{\text{DE}}
\rangle_{-1} \nonumber \\
&+ \frac{i}{\hbar} M(t) \big(\langle f^c_{1\uparrow} f^c_{1\downarrow} f^v_{2\uparrow}  \rangle - \langle f^c_{2\uparrow}
f^c_{1\uparrow} f^c_{1\downarrow} \rangle\big)
\nonumber \\
&+ \frac{i}{\hbar} ( \epsilon^c_{2\ua} - \epsilon^v_{2\ua}  -\hbar \omega_{L} + i \hbar \gamma_{\text{PD}}) \langle f^c_{1\uparrow}
f^c_{1\downarrow} p_{2\uparrow}  \rangle_{-1}
\end{align}
Finally, the dynamical equations for all higher correlations are given by
\begin{align}
d_t&\langle f^c_{1\downarrow} f^v_{1\downarrow} p_{2\uparrow}  \rangle_{-1}	= 
\frac{i}{\hbar} M(t) \big( \langle f^c_{1\downarrow} f^v_{2\uparrow} f^v_{1\downarrow}  \rangle
- \langle f^c_{2\uparrow} f^c_{1\downarrow} f^v_{1\downarrow}  \rangle\big)
\nonumber
\\
&+ \frac{i}{\hbar} ( \epsilon^c_{2\ua} - \epsilon^v_{2\ua}  -\hbar \omega_{L} + i \hbar \gamma_{\text{PD}}) \langle
f^c_{1\downarrow} f^v_{1\downarrow} p_{2\uparrow}  \rangle_{-1}\end{align}
\begin{align}
d_t&\langle f^c_{1\uparrow} f^v_{1\downarrow} p_{2\uparrow}  \rangle_{-1}	= 
\frac{i}{\hbar} M(t) \big( \langle f^c_{1\uparrow} f^v_{2\uparrow} f^v_{1\downarrow}  \rangle
- \langle f^c_{2\uparrow} f^c_{1\uparrow} f^v_{1\downarrow}  \rangle\big)
\nonumber
\\
&+ \frac{i}{\hbar} ( \epsilon^c_{2\ua} - \epsilon^v_{2\ua}  -\hbar \omega_{L} + i \hbar \gamma_{\text{PD}}) \langle
f^c_{1\uparrow} f^v_{1\downarrow} p_{2\uparrow}  \rangle_{-1}\end{align}
\begin{align}
d_t \langle f^c_{2\uparrow} f^c_{1\downarrow} f^v_{1\downarrow}  \rangle	= 
\hzwei M(t) \imag{\langle f^c_{1\downarrow} f^v_{1\downarrow} p_{2\uparrow} \rangle_{-1}}
\end{align}
\begin{align}
d_t \langle f^c_{2\uparrow} f^c_{1\uparrow} f^c_{1\downarrow}  \rangle	= 
\hzwei M(t) \imag{\langle f^c_{1\uparrow} f^c_{1\downarrow} p_{2\uparrow}  \rangle_{-1}}
\end{align}
\begin{align}
d_t \langle f^c_{2\uparrow} f^c_{1\uparrow} f^v_{1\downarrow}  \rangle	= 
\hzwei M(t) \imag{\langle f^c_{1\uparrow} f^v_{1\downarrow} p_{2\uparrow} \rangle_{-1}}
\end{align}
\begin{align}
d_t&\langle f^c_{2\uparrow} p_{\text{DE}}  \rangle_{-1}	=
- \frac{i}{\hbar} M(t) \langle f^v_{2\uparrow} p_{\text{II}}  \rangle
+ \frac{i}{\hbar} V \langle f^c_{1\uparrow} f^c_{1\downarrow} p_{2\uparrow} \rangle_{-1}
\nonumber
\\&+ \frac{i}{\hbar} ( \epsilon^c_{1\ua} + \epsilon^c_{1\da} - \epsilon^v_{1\da} - \epsilon^v_{2\ua}  -\hbar \omega_{L} + i
\gamma_{\text{PD}} \hbar) \langle f^c_{2\uparrow} p_{\text{DE}}  \rangle_{-1} \end{align}
\begin{align}
d_t \langle f^c_{1\uparrow} f^c_{1\downarrow} f^v_{2\uparrow}  \rangle	= 
- \hzwei M(t) \imag{\langle f^c_{1\uparrow} f^c_{1\downarrow} p_{2\uparrow}  \rangle_{-1}}
\nonumber
\\ 
+ \hzwei V \imag{ \langle f^v_{2\uparrow} p_{\text{II}} \rangle}
\end{align}
\begin{align}
d_t \langle f^c_{1\downarrow} f^v_{2\uparrow} f^v_{1\downarrow}  \rangle	= 
- \hzwei M(t) \imag{\langle f^c_{1\downarrow} f^v_{1\downarrow} p_{2\uparrow}  \rangle_{-1}}
\end{align}
\begin{align}
d_t \langle f^c_{1\uparrow} f^v_{2\uparrow} f^v_{1\downarrow}  \rangle	= 
- \hzwei M(t) \imag{ \langle f^c_{1\uparrow} f^v_{1\downarrow} p_{2\uparrow}  \rangle_{-1}}
\end{align}
\begin{align}
d_t&\langle f^v_{2\uparrow} p_{\text{II}}  \rangle	= 
- \frac{i}{\hbar} M(t) \langle f^c_{2\uparrow} p_{\text{DE}}  \rangle_{-1}
  \nonumber
\\ 
&+ \frac{i}{\hbar} V \big(\langle f^c_{2\uparrow} f^v_{2\uparrow} f^v_{1\downarrow} 
\rangle
- \langle f^c_{1\uparrow} f^c_{1\downarrow} f^v_{2\uparrow} \rangle\big)
\nonumber
\\
&+ \frac{i}{\hbar} ( \epsilon^c_{1\ua} + \epsilon^c_{1\da} - \epsilon^v_{1\da} - \epsilon^c_{2\ua} + i \gamma_{\text{PD}}
\hbar) \langle f^v_{2\uparrow} p_{\text{II}}  \rangle
\end{align}
\begin{align}
d_t \langle f^c_{2\uparrow} f^v_{2\uparrow} f^v_{1\downarrow}  \rangle	= 
- \hzwei V \imag{\langle f^v_{2\uparrow} p_{\text{II}} \rangle}
\end{align}

The higher electronic correlations which are not driven are not displayed here.

\subsection{Quantum-optical system}

Next, we show the photon-assisted correlations, again restricting to the impact ionization and Auger recombination processes
in the conduction band.

\begin{align}
d_t \langle \cdag{\textbf k2} \can{\textbf k1}  \rangle	= 
\frac{i}{\hbar} M_{2\uparrow}^{\textbf k}\langle p_{2\uparrow} \can{\textbf k1}  \rangle
\nonumber
\nonumber
- \frac{i}{\hbar} M_{2\uparrow}^{\textbf k *}\langle p_{2\uparrow}^\dagger \cdag{\textbf k2}  \rangle
\newline\\+ \frac{i}{\hbar} (\hbar \omega_{\textbf k2} - \hbar \omega_{\textbf k1} ) \langle \cdag{\textbf k2} \can{\textbf k1} 
\rangle\end{align}

\begin{align}
d_t&\langle p_{2\uparrow} \can{\textbf k1}  \rangle	= 
\nonumber
\frac{i}{\hbar} M(t) \langle f^v_{2\uparrow} \can{\textbf k1}  \rangle_{+1}
\nonumber
- \frac{i}{\hbar} M(t) \langle f^c_{2\uparrow} \can{\textbf k1}  \rangle_{+1}\\ 
\nonumber
&+ \frac{i}{\hbar} V \langle \pDE \can{\textbf k1} 
\rangle
\nonumber
- \frac{i}{\hbar} M_{2\uparrow}^{\textbf k *}\langle f^c_{2\uparrow}  \rangle
\nonumber
+ \frac{i}{\hbar} M_{2\uparrow}^{\textbf k *}\langle f^c_{2\uparrow} f^v_{2\uparrow}  \rangle
\newline\\&+ \frac{i}{\hbar} (\epsilon^c_{2\ua} - \epsilon^v_{2\ua} - \hbar \omega_{\textbf k1} + i \gamma_{\text{PD}} \hbar) \langle
p_{2\uparrow} \can{\textbf k1}  \rangle\end{align}

\begin{align}
d_t&\langle f^v_{2\uparrow} \can{\textbf k1}  \rangle_{+1}	= 
\nonumber
\frac{i}{\hbar} M(t) \big( \langle p_{2\uparrow} \can{\textbf k1}  \rangle
\nonumber
- \langle p_{2\uparrow}^\dagger \can{\textbf k1}  \rangle_{+2} \big)
\\
&- \frac{i}{\hbar} M_{2\uparrow}^{\textbf k *}\langle p_{2\uparrow}^\dagger  \rangle_{+1}
+ \frac{i}{\hbar} (- \hbar \omega_{\textbf k1}  + \omega_{L} ) \langle f^v_{2\uparrow} \can{\textbf k1} 
\rangle_{+1}\end{align}

\begin{align}
d_t&\langle f^c_{2\uparrow} \can{\textbf k1}  \rangle_{+1}	= 
\nonumber
\frac{i}{\hbar} M(t) \big( \langle p_{2\uparrow}^\dagger \can{\textbf k1}  \rangle_{+2}
- \langle p_{2\uparrow} \can{\textbf k1}  \rangle \big)\\
\nonumber
\nonumber
&+ \frac{i}{\hbar} V \langle \pII \can{\textbf k1} 
\rangle_{+1}
\nonumber
- \frac{i}{\hbar} V \langle \pII^\dagger \can{\textbf k1} 
\rangle_{+1}
\newline\\&+ \frac{i}{\hbar} (- \hbar \omega_{\textbf k1}  + \omega_{L} ) \langle f^c_{2\uparrow} \can{\textbf k1} 
\rangle_{+1}\end{align}

\begin{align}
d_t&\langle p_{2\uparrow}^\dagger \can{\textbf k1}  \rangle_{+2}	= 
- \frac{i}{\hbar} V \langle \pDE^\dagger \can{\textbf k1} \rangle_{+2}
\\ 
\nonumber
&+ \frac{i}{\hbar} M(t) \langle f^c_{2\uparrow} \can{\textbf k1}  \rangle_{+1}
\nonumber
- \frac{i}{\hbar} M(t) \langle f^v_{2\uparrow} \can{\textbf k1}  \rangle_{+1}
\nonumber
\newline\\&+ \frac{i}{\hbar} (\epsilon^v_{2\ua} - \epsilon^c_{2\ua} - \hbar \omega_{\textbf k1}  +2 \omega_{L} + i \gamma_{\text{PD}}
\hbar) \langle p_{2\uparrow}^\dagger \can{\textbf k1}  \rangle_{+2} \nonumber\end{align}

\begin{align}
d_t&\langle \pDE \can{\textbf k1}  \rangle	= 
- \frac{i}{\hbar} M(t) \langle \pII \can{\textbf k1}  \rangle_{+1}\\ 
\nonumber
&+ \frac{i}{\hbar} V \langle f^v_{1\downarrow} p_{2\uparrow} \can{\textbf k1} 
\rangle \nonumber
- \frac{i}{\hbar} M_{2\uparrow}^{\textbf k *}\langle \pII  \rangle
\newline\\&+ \frac{i}{\hbar} (\epsilon^c_{1\ua} + \epsilon^c_{1\da} - \epsilon^v_{1\da} - \epsilon^v_{2\ua} - \hbar \omega_{\textbf
k1} + i \gamma_{\text{PD}} \hbar) \langle \pDE \can{\textbf k1}  \rangle \nonumber\end{align}

\begin{align}
d_t&\langle \pII \can{\textbf k1}  \rangle_{+1}	= 
- \frac{i}{\hbar} M(t) \langle \pDE \can{\textbf k1}  \rangle\\ 
\nonumber
&+ \frac{i}{\hbar} V \big( \langle f^c_{2\uparrow} f^v_{1\downarrow} \can{\textbf k1} 
\rangle_{+1}
- \langle f^c_{1\uparrow} f^c_{1\downarrow} \can{\textbf k1} 
\rangle_{+1} \big)
\nonumber
\newline\\&+ \frac{i}{\hbar} (\epsilon^c_{1\ua} + \epsilon^c_{1\da} - \epsilon^v_{1\da} - \epsilon^c_{2\ua} - \hbar \omega_{\textbf
k1}  + \omega_{L} + i \gamma_{\text{PD}} \hbar) \langle \pII \can{\textbf k1} 
\rangle_{+1} \nonumber\end{align}

\begin{align}
d_t&\langle \pII^\dagger \can{\textbf k1}  \rangle_{+1}	=
\frac{i}{\hbar} M(t) \langle \pDE^\dagger \can{\textbf k1}  \rangle_{+2}\\ 
\nonumber
&+ \frac{i}{\hbar} V \big( \langle f^c_{1\uparrow} f^c_{1\downarrow} \can{\textbf k1} 
\rangle_{+1}
\nonumber
- \langle f^c_{2\uparrow} f^v_{1\downarrow} \can{\textbf k1} 
\rangle_{+1} \big)
\nonumber
\newline\\&+ \frac{i}{\hbar} (\epsilon^c_{2\ua} + \epsilon^v_{1\da} - \epsilon^c_{1\da} - \epsilon^c_{1\ua} - \hbar \omega_{\textbf
k1}  + \omega_{L} + i \gamma_{\text{PD}} \hbar) \langle \pII^\dagger \can{\textbf k1} 
\rangle_{+1} \nonumber\end{align}

\begin{align}
d_t&\langle f^v_{1\downarrow} p_{2\uparrow} \can{\textbf k1}  \rangle	= 
\frac{i}{\hbar} M(t) \big( \langle f^v_{2\uparrow} f^v_{1\downarrow} \can{\textbf k1}  \rangle_{+1}
\nonumber
- \langle f^c_{2\uparrow} f^v_{1\downarrow} \can{\textbf k1}  \rangle_{+1} \big)\\ 
\nonumber
&+ \frac{i}{\hbar} V \langle \pDE \can{\textbf k1} 
\rangle 
\nonumber
- \frac{i}{\hbar} M_{2\uparrow}^{\textbf k *}\langle f^c_{2\uparrow} f^v_{1\downarrow}  \rangle
\nonumber
\newline\\&+ \frac{i}{\hbar} (\epsilon^c_{2\ua} - \epsilon^v_{2\ua} - \hbar \omega_{\textbf k1} + i \gamma_{\text{PD}} \hbar) \langle
f^v_{1\downarrow} p_{2\uparrow} \can{\textbf k1}  \rangle \end{align}

\begin{align}
d_t&\langle \pDE^\dagger \can{\textbf k1}  \rangle_{+2}	= 
\nonumber
\frac{i}{\hbar} M(t) \langle \pII^\dagger \can{\textbf k1}  \rangle_{+1}
\nonumber
- \frac{i}{\hbar} V \langle f^v_{1\downarrow} p_{2\uparrow}^\dagger \can{\textbf k1} 
\rangle_{+2}
\nonumber
\newline\\&+ \frac{i}{\hbar} (\epsilon^v_{2\ua} + \epsilon^v_{1\da} - \epsilon^c_{1\da} - \epsilon^c_{1\ua} - \hbar \omega_{\textbf
k1}  +2 \omega_{L} + i \gamma_{\text{PD}} \hbar) \langle \pDE^\dagger \can{\textbf k1} 
\rangle_{+2}\end{align}

\begin{align}
d_t&\langle f^c_{2\uparrow} f^v_{1\downarrow} \can{\textbf k1}  \rangle_{+1}	= 
\nonumber
\frac{i}{\hbar} M(t) \big( \langle f^v_{1\downarrow} p_{2\uparrow}^\dagger \can{\textbf k1}  \rangle_{+2}
\nonumber
- \langle f^v_{1\downarrow} p_{2\uparrow} \can{\textbf k1}  \rangle \big)\\ 
\nonumber
&+ \frac{i}{\hbar} V \big( \langle \pII \can{\textbf k1} 
\rangle_{+1}
- \langle \pII^\dagger \can{\textbf k1} \rangle_{+1} \big)
\nonumber
\newline\\&+ \frac{i}{\hbar} (- \hbar \omega_{\textbf k1}  + \omega_{L} ) \langle f^c_{2\uparrow} f^v_{1\downarrow}
\can{\textbf k1}  \rangle_{+1}\end{align}

\begin{align}
d_t&\langle f^c_{1\uparrow} f^c_{1\downarrow} \can{\textbf k1}  \rangle_{+1}	= 
\nonumber
\nonumber
+ \frac{i}{\hbar} V \big( \langle \pII^\dagger \can{\textbf k1} 
\rangle_{+1}
- \langle \pII \can{\textbf k1} 
\rangle_{+1} \big)
\nonumber
\newline\\&+ \frac{i}{\hbar} (- \hbar \omega_{\textbf k1}  + \omega_{L} ) \langle f^c_{1\uparrow} f^c_{1\downarrow}
\can{\textbf k1}  \rangle_{+1}\end{align}

\begin{align}
d_t&\langle f^v_{2\uparrow} f^v_{1\downarrow} \can{\textbf k1}  \rangle_{+1}	= 
\nonumber
- \frac{i}{\hbar} M_{2\uparrow}^{\textbf k *}\langle f^v_{1\downarrow} p_{2\uparrow}^\dagger  \rangle_{+1}\\
&+ \frac{i}{\hbar} M(t) \big( \langle f^v_{1\downarrow} p_{2\uparrow} \can{\textbf k1}  \rangle
\nonumber
-  \langle f^v_{1\downarrow} p_{2\uparrow}^\dagger \can{\textbf k1}  \rangle_{+2} \big)
\nonumber
\newline\\&+ \frac{i}{\hbar} (- \hbar \omega_{\textbf k1}  + \omega_{L} ) \langle f^v_{2\uparrow} f^v_{1\downarrow}
\can{\textbf k1}  \rangle_{+1}\end{align}

\begin{align}
d_t&\langle f^v_{1\downarrow} p_{2\uparrow}^\dagger \can{\textbf k1}  \rangle_{+2}	= 
- \frac{i}{\hbar} V \langle \pDE^\dagger \can{\textbf k1} 
\rangle_{+2}
\\
&+ \frac{i}{\hbar} M(t) \big( \langle f^c_{2\uparrow} f^v_{1\downarrow} \can{\textbf k1}  \rangle_{+1}
\nonumber
- \langle f^v_{2\uparrow} f^v_{1\downarrow} \can{\textbf k1}  \rangle_{+1} \big)
\nonumber
\newline\\&+ \frac{i}{\hbar} (\epsilon^v_{2\ua} - \epsilon^c_{2\ua} - \hbar \omega_{\textbf k1}  +2 \omega_{L} + i \gamma_{\text{PD}}
\hbar) \langle f^v_{1\downarrow} p_{2\uparrow}^\dagger \can{\textbf k1}  \rangle_{+2} \nonumber\end{align}

\end{document}